\documentstyle[preprint,aps,epsf,rotate]{revtex}
\draft 
\tightenlines


\begin{document}

\title{`Generalized des Cloizeaux' exponent for self-avoiding walks on 
        the incipient percolation cluster}
       
\author{Anke~Ordemann$^1$,
                Markus~Porto$^2$, H.~Eduardo~Roman$^{3}$, 
                and Shlomo~Havlin$^{4}$}

\address{$^1$Institut~f\"ur~Theoretische~Physik~III,
         Justus--Liebig--Universit\"at~Giessen,
         Heinrich--Buff--Ring~16,
         35392~Giessen, Germany   \\
         $^2$Max-Planck-Institut~f\"ur~Physik~komplexer~Systeme,
         N\"othnitzer~Str.~38, 01187~Dresden, Germany \\
         $^3$I.N.F.N., Sezione~di~Milano, Via~Celoria~16, 20133~Milano, Italy \\
         $^4$Minerva~Center~and~Department~of~Physics, Bar-Ilan~University,
         52900~Ramat-Gan, Israel}

\date{received \today}

\maketitle

\begin{abstract} 
We study the asymptotic shape of self-avoiding random walks (SAW) on the
backbone of the  incipient percolation cluster in $d$-dimensional lattices
analytically. It is generally accepted  that  the configurational averaged
probability distribution function  $\left< P_{\mathrm{B}}(r,N) \right>$ for the
end-to-end distance $r$ of an $N$ step SAW behaves as  a power law for $r\to
0$. In this work, we determine the corresponding exponent using scaling
arguments, and show that our suggested `generalized des Cloizeaux' expression
for the exponent  is in  excellent agreement with exact enumeration results in
two and three dimensions. 
\end{abstract}

\bigskip


Linear polymers have a broad spectrum of applications in such different fields
as biology, chemistry, and physics, see
e.g.~\cite{Dullien:1979,Cantor/Schimmel:1980,Andrews:1986}, and theoretical
work has been extensively documented in the literature
\cite{deGennes79,Doi/Edwards:1986,desCloizeaux/Jannik:1990}. More recently,
much interest has been drawn to the effects that environmental disorder may
have on the configurational properties of such linear structures
\cite{Nakanishi:1994,Barat/Chakrabarti:1995,Baumgaertner:1995}.  As a
paradigmatic model of linear polymers, self-avoiding random walks  (SAW) have 
captured the attention of researchers for decades and are currently
extensively 
studied\cite{deGennes79,Doi/Edwards:1986,desCloizeaux/Jannik:1990}. 
Structural disorder is generally modelled by percolation
\cite{Stauffer/Aharony:1992,Bunde/Havlin:1996}, and interesting effects are
expected at the percolation threshold\cite{Barat/Chakrabarti:1995}. Since
asymptotically long SAW can only exist on the backbone of the cluster, in
which all dangling ends have been eliminated, one may study SAW directly on
the  backbone. For relatively short chains, one may still expect that the
statistical properties of SAW do not depend on whether they are studied on the
backbone alone  or on the corresponding percolation cluster. In what follows,
we adopt this point of view and study SAW directly on the backbone of the
cluster.

On percolation clusters, one can distinguish between two different metrics, the
usual Pythagorian one, or $r$-space metric, and the topological one, or
$\ell$-space metric, which have been extensively discussed in the literature
\cite{Bunde/Havlin:1996,Grassberger:1992,Grassberger:1999b}. The corresponding 
length scales are related by $\ell\sim r^{d_{\rm min}}$, where $d_{\rm min}$ 
is the fractal dimension of the shortest path \cite{Bunde/Havlin:1996}.

The scaling behavior of SAW is characterized by the mean end--to--end distance
after $N$ steps. On structurally disordered systems such as percolation
clusters, two  types  of averages need to be performed. One first averages over
all SAW  configurations of $N$ steps on a single backbone starting from the
same origin, as obtained e.g. by exact enumeration 
techniques~\cite{Grassberger:1982}, followed by a configurational average over
many backbones. In $r$-space, the former average is denoted as $\bar{r^2}(N)$,
and the second one by $\langle \bar{r^2}(N) \rangle \sim N^{2 \nu_ r}$, which
defines the exponent $\nu_r$. Correspondingly, in $\ell$-space one has, 
$\langle \bar{\ell}(N) \rangle \sim N^{\nu_{\ell}}$, where
$\nu_\ell=\nu_r~d_{\rm min}$.

More generally, one is interested in the configurational averaged probability 
distribution function (PDF) for the end--to--end distance on the backbone. In
$r $-space, it is denoted as $\left< P_{\mathrm{B}}(r,N) \right>$, and is
expected to obey the scaling  form 
\begin{equation}\label{PrN} 
\left< P_{\mathrm{B}}(r,N) \right> = {1\over N^{\nu_r d_{\rm
f}^{\mathrm{B}}}}~f_r(r/N^{\nu_r})  \quad {\rm for} \quad N \gg 1 \quad , 
\end{equation} 
which is normalized according to  $\int_0^{\infty} {\rm d}r~r^{d_{\rm
f}^{\mathrm{B}}-1} \left< P_{\mathrm{B}}(r,N) \right>=1$, where $d_{\rm
f}^{\mathrm{B}}$ is the fractal  dimension of the backbone. The scaling
function $f_r(x)$ behaves as  $f_r(x) \sim x^{g_1^r}$, for $x \ll 1$, and in
this limit Eq.~(\ref{PrN}) becomes 
\begin{equation}\label{Prto0N} 
\left< P_{\mathrm{B}}(r,N) \right> \sim {1\over N^{\nu_r d_{\rm
f}^{\mathrm{B}}} }~ \left( {r \over N^{\nu_r}} \right)^{g_1^r} \quad {\rm for}
\quad r\ll N^{\nu_r} \quad , 
\end{equation} 
where $g_1^r$ is the scaling exponent to which we draw our attention here. In
$\ell$-space, the PDF obeys  $\left< P_{\mathrm{B}}(\ell,N) \right> =
N^{-\nu_{\ell} d_{\ell}^{\mathrm{B}}}~ f_{\ell}(\ell/N^{\nu_\ell})$  and is
normalized according to $\int_0^{\infty} {\rm d}\ell~{\ell}^{d_{\rm
\ell}^{\mathrm{B}}-1} \left< P_{\mathrm{B}}(\ell,N) \right>=1$, with 
$d_{\ell}^{\mathrm{B}}$ being the fractal dimension of the backbone in
$\ell$-space, $d_{\ell}^{\mathrm{B}} = d_{\mathrm{f}}^{\mathrm{B}} /
d_{\mathrm{min}}$. The corresponding exponent in the case $\ell\ll
N^{\nu_{\ell}}$ is denoted by  $g_1^{\ell}$ and is related to $g_1^r$ by
$g_1^{\ell}=g_1^r/d_{\rm min}$ (see e.g.\ 
\cite{Roman/Ordemann/Porto/Bunde/Havlin:1998,Ordemann/Porto/Roman/Havlin/Bunde:2000}).
The interest in studying  the PDF $\left< P_{\mathrm{B}}(\ell,N) \right>$ is
because  fluctuations are definitely smaller in $\ell$-space than in
$r$-space, and more  accurate  results for $g_1^r$ can be derived by determining 
$g_1^{\ell}$ and afterwards using the
relation $g_1^r=g_1^{\ell} d_{\rm min}$.

In the case of regular lattices ($g_1^{\ell}=g_1^r=g_1$), des
Cloizeaux~\cite{desCloizeaux:1974} showed that $g_1=(\gamma-1)/\nu_{\rm F}$,
where  $\nu_{\rm F}\cong 3/(d+2)$ is the Flory exponent \cite{Flory:1949} and
$\gamma$  describes  the total number $C_N$ of SAW configurations of $N$ steps,
\begin{equation} \label{CN_reg} 
C_{N} \cong \mu^N \, N^{\gamma -1} \quad {\rm for} \quad N \gg 1 \quad , 
\end{equation} 
where $\mu$ is the effective coordination number of the lattice (see
e.g.\ \cite{desCloizeaux/Jannik:1990}). Numerical results for SAW on the
backbone of percolation clusters at criticality in two dimensions seem to be
roughly consistent with the form $g_1^r=(\gamma_1-1)/\nu_r$
~\cite{Roman/Ordemann/Porto/Bunde/Havlin:1998},  where $\gamma_1$ is the
corresponding enhancement exponent on the backbone. More recent results in
three dimensions \cite{Ordemann/Porto/Roman/Havlin/Bunde:2000} clearly disprove
this relation, so the question remains whether a generalization of the des
Cloizeaux expression can still provide an accurate framework for estimating the
exponent $g_1^r$ analytically. The goal of the present paper is to provide such
a generalization of the des Cloizeaux relation valid in all dimensions.

To this end, we follow de Gennes~\cite{deGennes79} and study the probability 
$\left< P_{\mathrm{B}}(a,N) \right>$ that the $N$-th step of the SAW is on a
backbone site  located at a lattice distance $a=1$ from its starting point at
$r=0$. This probability can be written as
\begin{equation}\label{P_N,B1}
\langle P_{\mathrm{B}}(a=1,N) \rangle  \sim 
\frac{ { \langle C_{N,{\mathrm{B}}}}(a=1)\rangle  }
     { { \langle C_{N,{\mathrm{B}}} \rangle } }
\end{equation}
where $\langle C_{N,{\mathrm{B}}}(a=1)\rangle $ represents the
(configurational) average  number of SAW that, after $N$ steps, arrive at a
distance $a=1$ from the origin, and  $\langle C_{N,{\mathrm{B}}}\rangle $
denotes the total number of SAW of $N$ steps.  The latter is the
configurational average analog of Eq.~(\ref{CN_reg}), with $\mu$ and  $\gamma$
replaced by the exponents $\mu_1$ and $\gamma_1$. The index 1 reminds us of the underlying 
multifractal behaviour of these two exponents for percolation systems 
\cite{Ordemann/Porto/Roman/Havlin/Bunde:2000} (see also
\cite{Grassberger:1993}). We assume the former to be
\begin{equation}\label{C_N,B1}
\langle C_{N,{\mathrm{B}}}(a=1) \rangle = \mu_{1}^N \, 
\left( {a\over R} \right)^{d_{\mathrm{f}}^{\mathrm{B}}} \, F_{\mathrm{B}}
\end{equation}
where $R \equiv \langle \bar{r^2}(N) \rangle^{1/2} \sim N^{\nu_r}$, and the
factor  $F_{\mathrm{B}} \sim R^{-\kappa} \le 1$ represents the additional
difficulty of the $N$ step SAW to return close to its starting point due to
topological constraints caused by the statistical nature of the embedding
structure.
For regular lattices, this factor reduces to unity since in those cases
there is no distinction between cluster  and lattice sites, and the exponent
$d_{\mathrm{f}}^{\mathrm{B}}$ is replaced by the  spatial dimensionality $d$
of the lattice \cite{deGennes79}. For deterministic fractals such as
Sierpinski gaskets, the factor $F_{\mathrm{B}}$ is also expected to be unity, 
since in those cases the fractal structure embedding the SAW is also 
unique~\cite{Ordemannetal:2000}. 

The exponent $\kappa$ can be determined by considering first the probability
that an arbitrary 
site belongs to the infinite cluster, $P_{\infty}$. Near the percolation threshold 
$p_{\mathrm{c}}$, $P_{\infty}$ behaves as a function of the concentration 
$p$ ($\ge p_{\mathrm{c}}$) of occupied sites as 
\cite{Stauffer/Aharony:1992,Bunde/Havlin:1996}
\begin{equation}\label{P_infty}
P_{\infty} \sim  {(p-p_{\mathrm{c}})}^{\beta} \sim \xi^{-\beta/\nu} 
\end{equation}
where the correlation length $\xi$, diverging as $\xi\sim{|p-p_{\mathrm{c}}|}^{-\nu}$ 
near $p_{\mathrm{c}}$, is a measure of the linear size of the finite clusters in the 
system (or similarly, the mean distance between two sites on the same finite
cluster), 
and $\nu$ is the correlation length exponent. Eq.\ (\ref{P_infty}) suggests that the
probability $P_{\infty}(L)$, to find a site belonging to the incipient cluster
within a 
distance $L$ from a given cluster site (and consequently at least one path connecting
these two sites), scales as $P_{\infty}(L) \sim L^{-\beta /\nu}$
for $L < \xi$\cite{Bunde/Havlin:1996}. 
Identifying the length scale $L$ with the mean size $R$ of the SAW, $L\sim R$, we obtain 
$F_{\mathrm{B}} \sim R^{-\beta /\nu}$, i.e. $\kappa = \beta/\nu$.
Accepted values for the critical exponents reported so 
far~\cite{Stauffer/Aharony:1992,Bunde/Havlin:1996,Grassberger:1992,Grassberger:1999b,Nienhuis:1982,Strenski/Bradley/Debierre:1991,Essam:1980,Grassberger:1999a,Porto/Bunde/Havlin/Roman:1997}
 are summarized in Table~\ref{table1}.

Using Eqs.~(\ref{P_N,B1}), (\ref{C_N,B1}) and the above result for $F_{\mathrm{B }}$, we find 
\begin{equation}\label{PBa1} 
\langle P_{\mathrm{B}}(a=1,N) \rangle  \sim 
\frac{\mu_{1}^N \,{(a/R)}^{d_{\mathrm{f}}^{\mathrm{B}}} \, 
F_{\mathrm{B}}}{\mu_{1}^N \, {N}^{\gamma_1 -1}} \sim { 1\over N^{\nu_r
d_{\mathrm{f}}^{\mathrm{B}}} }   \left( {1\over N^{\nu_r}}
\right)^{(\gamma_1-1)/\nu_r+\beta/\nu},
\end{equation} 
and comparison with Eq.~(\ref{Prto0N}) yields 
\begin{equation}\label{gendeCloi} 
g_1^r = { \gamma_1 - 1 \over \nu_r} + { \beta \over \nu }
\end{equation} 
which is denoted as the generalized des Cloizeaux relation.

Numerical calculations based on exact enumeration techniques in two and three
dimensions (taken from Ref.\ \cite{Ordemann/Porto/Roman/Havlin/Bunde:2000})
are reported in Fig.~\ref{fig:exenud2d3} for $\left< P_{\mathrm{B}}(\ell,N)
\right>$~\cite{note}.  The corresponding numerical values for $g_1^{\ell}$ and $g_1^r$ are
summarized in  Table~\ref{table2}. The suggested relation
Eq.~(\ref{gendeCloi}) is in excellent agreement  with the numerical data.
Additionally, it should be noted  that Eq.~(\ref{gendeCloi}) yields the
correct  value expected when $d \ge 6$, i.e. $g_1^r = 2$. Indeed, when $d \ge
6$ the backbone of  the incipient percolation cluster becomes topologically
one dimensional, since loops are  irrelevant on large length scales, and one
simply has $\gamma_1=1$. Thus, $g_1^r $ is solely  determined by the second
term in Eq.~(\ref{gendeCloi}). The value $g_1^r=2$ for  $d\ge 6$  can be
understood by noting that in this case  $F_{\mathrm{B}} \sim
{(a/R)}^{d^{\mathrm{B}}_{\mathrm{f}}}$ with $d^{\mathrm{B}}_{\mathrm{f}} = 2$
for $d \ge 6$, since SAW  are equivalent to the embedding backbone and the
probability to return close to the starting point just decreases as $R^{-2}$,
i.e. $g_1^r = 2$. The good agreement between  Eq.~(\ref{gendeCloi}) and the
numerical results supports the ansatz made in  Eq.~(\ref{C_N,B1}).


Financial support from the German--Israeli Foundation (GIF),
the Minerva Center for Mesoscopics, Fractals, and Neural
Networks and the Deutsche Forschungsgemeinschaft is gratefully acknowledged.


\narrowtext
\begin{figure}[t]
\caption{Scaling plots of the PDF in $\ell$-space on the backbone of the
incipient percolation cluster in  {\bf (a)},{\bf (b)} $d=2$  and  {\bf
(c)},{\bf(d)} $d=3$, from the exact enumeration results of Ref.\
\protect\cite{Ordemann/Porto/Roman/Havlin/Bunde:2000}. In {\bf (a)} and {\bf
(c)}, the whole PDF is shown, plotted as  $\ell^{d_{\ell}^{\mathrm{B}}} \left<
{P}_{\mathrm{B}}(\ell,N) \right> $ versus $\ell/N^{\nu_{\ell}}$, for $N=39$
(full circles) and $40$ (full squares), averaged over $5 \cdot 10^3$ backbone
configurations.  
In {\bf (b)} and {\bf (d)},  the region with $\ell \ll
N^{\nu_{\ell}}$ is shown in more detail, plotted as
$\ell^{d_{\ell}^{\mathrm{B}}}~\left< P_{\mathrm{B}}(\ell,N) \right> 
/{(\ell/N^{\nu_\ell})}^{d_{\ell}^{\mathrm{B}}} =
{(N^{\nu_\ell})}^{d_{\ell}^{\mathrm{B}}} \, \langle {P}_{\mathrm{B}}(\ell,N)
\rangle $ versus $\ell/N^{\nu_{\ell}}$~\protect\cite{note}. Case $d=2$: In {\bf
(a)} the continuous line corresponds to the  case  $\ell\gg N^{\nu_{\ell}}$ and
is not further discussed here 
(see~\protect\cite{Ordemann/Porto/Roman/Havlin/Bunde:2000} for details). For
$\ell \ll N^{\nu_{\ell}}$ the dashed line in {\bf (b)} has the slope $0.49 \pm
0.05$ and is a fit with the ansatz  ${(N^{\nu_\ell})}^{d_{\ell}^{\mathrm{B}}}
\, \left< P_{\mathrm{B}}(\ell,N) \right>   \sim (\ell/N^{\nu_\ell})^{g_1^{\ell}
}$   for $\ell \ll N^{\nu_{\ell}}$ (cf.\ Eq.~(\protect\ref{Prto0N}) and text),
in very good agreement with the  prediction $g_1^{\ell} = (\gamma_1
-1)/\nu_{\ell} + \beta/(\nu d_{\mathrm{min}})\protect\cong 0.474$  (derived
from Eq.~(\protect\ref{gendeCloi}) using the relation $g_1^{\ell}=g_1^r/d_{\rm
min}$).  The dotted line is for illustration only and has the slope $0.382$,
resulting from the  relation  $  (\gamma_1 -1) /\nu_{\ell}\protect\cong 0.382$,
in disagreement with the numerical results. Case $d=3$: Similarly to  {\bf
(b)}, in {\bf (d)} the dashed line has the slope $0.67 \pm 0.05$, in excellent
agreement with the  prediction  $g_1^{\ell} = (\gamma_1 -1)/\nu_{\ell} +
\beta/(\nu d_{\mathrm{min}})\protect\cong 0.666$. The dotted line has the slope
$0.32$, resulting from $ (\gamma_1 -1)/\nu_{\ell}\protect\cong 0.32$,  clearly
revealing the inadequacy of this relation to fit the numerical results.} 
\label{fig:exenud2d3}
\end{figure}\vfill


\widetext
\begin{table}[b]
\begin{center}
\begin{tabular}{c||r@{}l|r@{}l|r@{}l}
 & \multicolumn{2}{c}{$d=2$} & \multicolumn{2}{c}{$d=3$} & \multicolumn{2}{c}{$d \ge 6$} \\
\hline & & & & \\[-4.0mm] 
$\beta$  & $5/36$ & \raisebox{1ex}{{\scriptsize a}}\,\ignorespaces   & $0.417 \pm 0.003$ & \raisebox{1ex}{{\scriptsize b}}\,\ignorespaces   & $1$  & \raisebox{1ex}{{\scriptsize c}}\,\ignorespaces  \\
\hline
$\nu$    & $4/3$ & \raisebox{1ex}{{\scriptsize a}}\,\ignorespaces     & 
$0.875 \pm 0.008$ & \raisebox{1ex}{{\scriptsize b}}\,\ignorespaces   & $1/2$ & \raisebox{1ex}{{\scriptsize c}}\,\ignorespaces  \\
\hline
$d_{\mathrm{f}}$ & $91/48$ & \raisebox{1ex}{{\scriptsize a}}\,\ignorespaces  & $2.524\pm0.008$ & \raisebox{1ex}{{\scriptsize b}}\,\ignorespaces & $4$ & \raisebox{1ex}{{\scriptsize d}}\,\ignorespaces \\
\hline
$d_{\mathrm{f}}^{\mathrm{B}}$ & $1.6432 \pm 0.0008$ & \raisebox{1ex}{{\scriptsize e}}\,\ignorespaces  & $1.87\pm0.03$ & \raisebox{1ex}{{\scriptsize f}}\,\ignorespaces & $2$& \raisebox{1ex}{{\scriptsize d}}\,\ignorespaces  \\
\hline
$d_{\mathrm{min}}$                        & $1.1306 \pm0.0003$ & \raisebox{1ex}{{\scriptsize g}}\,\ignorespaces  & $1.374\pm0.004$& \raisebox{1ex}{{\scriptsize h}}\,\ignorespaces  & $2$ & \raisebox{1ex}{{\scriptsize d}}\,\ignorespaces \\
\hline
$d_{{\ell}}^{\mathrm{B}}$ & $1.446 \pm 0.001$ & \raisebox{1ex}{{\scriptsize i}}\,\ignorespaces  & $1.36\pm0.02$ & \raisebox{1ex}{{\scriptsize f}}\,\ignorespaces
 & $1$ & \raisebox{1ex}{{\scriptsize d}}\,\ignorespaces \\
\end{tabular}
\end{center}
\caption{Critical exponents and fractal dimensions for the incipient percolation
cluster, for spatial dimensions $d=2$, 3, and $d\ge 6$, from
\protect\raisebox{1ex}{{\scriptsize a}}\,\ignorespaces %
exact results Ref.~\protect\cite{Nienhuis:1982},
\protect\raisebox{1ex}{{\scriptsize b}}\,\ignorespaces %
Monte Carlo simulations Ref.~\protect\cite{Strenski/Bradley/Debierre:1991},
\protect\raisebox{1ex}{{\scriptsize c}}\,\ignorespaces %
exact results Ref.~\protect\cite{Essam:1980},
\protect\raisebox{1ex}{{\scriptsize d}}\,\ignorespaces %
Refs.~\protect\cite{Stauffer/Aharony:1992,Bunde/Havlin:1996},
\protect\raisebox{1ex}{{\scriptsize e}}\,\ignorespaces %
Monte Carlo simulations Ref.~\protect\cite{Grassberger:1999a},
\protect\raisebox{1ex}{{\scriptsize f}}\,\ignorespaces %
Monte Carlo simulations Ref.~\protect\cite{Porto/Bunde/Havlin/Roman:1997},
\protect\raisebox{1ex}{{\scriptsize g}}\,\ignorespaces %
Monte Carlo simulations Ref.~\protect\cite{Grassberger:1999b},
\protect\raisebox{1ex}{{\scriptsize h}}\,\ignorespaces %
Monte Carlo simulations Ref.~\protect\cite{Grassberger:1992},
and \protect\raisebox{1ex}{{\scriptsize i}}\,\ignorespaces %
the relation $d_{{\ell}}^{\mathrm{B}} = d_{\mathrm{f}}^{\mathrm{B}} /
d_{\mathrm{min}}$. Here, $d_{\mathrm{f}}=d-\beta/\nu$ is the fractal
dimension of percolation clusters.
}
\label{table1}
\end{table}

\widetext
\begin{table}[b]
\begin{center}
\begin{tabular}{c|c|c|c}
              &     $d=2$         &       $d=3$       &  $d \ge 6$   \\
\hline
$\nu_r$       & $0.787 \pm 0.010$ & $0.662 \pm 0.006$ &    $1/2$     \\
\hline
$\gamma_1$    & $1.34  \pm 0.05$  & $1.29  \pm 0.05$  &    $ 1 $     \\
\hline
$g_1^r$       & $0.56  \pm 0.10$  & $0.90  \pm 0.10$  &    $ 2 $     \\
\hline 
$g_1^{\ell}$  & $0.49  \pm 0.05$  & $0.67  \pm 0.05$  &    $ 1 $     \\
\hline
$g_1^r=g_1^{\ell}\, d_{\mathrm{min}}$ & $0.55\pm 0.06$ & $0.92 \pm 0.08$ & $2$ 
\\
\hline
$g_1^r$ (present conjecture) & $0.54 \pm 0.07$ & $0.916 \pm 0.080$ &    $ 2 $     \\
\end{tabular}
\end{center}
\caption{Results for SAW on the backbone of critical percolation clusters (from
Ref.~\protect\cite{Ordemann/Porto/Roman/Havlin/Bunde:2000}).  Note that the value
of $g_1^{\mathrm{\ell}}$  slightly differs from the value given in
Ref.~\protect\cite{Ordemann/Porto/Roman/Havlin/Bunde:2000}, as we determined it
more accurately here by plotting $\left< P_{\mathrm{B}}(\ell,N) \right>$ as shown
in Fig.~\ref{fig:exenud2d3}.  The last line displays the present theoretical
results from the generalized des Cloizeaux  relation
$g_1^r=(\gamma_1-1)/\nu_{r}+\beta/\nu$, Eq.~(\protect\ref{gendeCloi}).}
\label{table2}
\end{table}

\end{document}